\documentclass[twoside,fleqn]{article}
\usepackage{latexsym,espcrc2}

\newcommand{\AmS}{{\protect\the\textfont2
  A\kern-.1667em\lower.5ex\hbox{M}\kern-.125emS}}

\def\spose#1{\hbox to 0pt{#1\hss}}
\def\ltapprox{\mathrel{\spose{\lower 3pt\hbox{$\mathchar"218$}}
 \raise 2.0pt\hbox{$\mathchar"13C$}}}
\def\gtapprox{\mathrel{\spose{\lower 3pt\hbox{$\mathchar"218$}}
 \raise 2.0pt\hbox{$\mathchar"13E$}}}

\title{Monte Carlo results for three-dimensional self-avoiding walks}

\author{S. Caracciolo\address{Scuola Normale Superiore and INFN,
        Sezione di Pisa, I-56100 Pisa, ITALY},
        M. S. Causo\address{Dipartimento di Fisica and INFN, Universit\`a
        degli Studi di Lecce, I-73100 Lecce, ITALY}\thanks{Speaker at the
        conference.}
        \ and
        A. Pelissetto\address{Dipartimento di Fisica and INFN,
        Universit\`a degli Studi di Pisa,
        I-56100 Pisa, ITALY}}

\begin{document}
\begin{abstract}
We discuss possible sources of systematic errors in the computation of 
critical exponents  by renormalization-group methods, 
extrapolations from exact enumerations and Monte
Carlo simulations. A {\em careful} Monte Carlo determination of the 
susceptibility exponent
$\gamma$ for three-dimensional self-avoiding walks has been used to test 
the claimed accuracy of the various methods.
\end{abstract}

\maketitle

\section{Possible sources of systematic errors in the computation of critical
exponents}

The major problem in a high-precision determination of critical parameters is 
the presence of corrections to scaling. Indeed they are {\em a priori} unknown
and cause a systematic error which is usually difficult to detect. The 
problem appears in extrapolations from exact enumerations, where 
one is forced to use higher-order differential approximants which 
however give stable results only for long series.
The problem arises in Monte
Carlo (MC) simulations as well. Because the correlation length is
necessarily much smaller than the lattice size $L$ to avoid finite-size
effects, the simulation is done in a region of temperatures
so far from the critical point that neglecting the unknown
corrections to scaling can deeply affect the final estimates. A
clear example of this effect can be found in the history of the estimates 
of the
critical exponent $\nu$ for three-dimensional SAWs: the first MC simulations
\cite{Watts} with quite short walks suggested an exponent in agreement 
with the
Flory theory $\nu=3/5$, while a recent simulation \cite{LMS}
with very long walks (corresponding to a 
correlation length of  $\xi(\beta) \approx 340$) gave the
much lower estimate $\nu = 0.5877 \pm 0.0006$
which is in good agreement with renormalization-group (RG) estimates. 
It is clear that only a careful analysis of the Monte Carlo data 
can prevent from underestimating the error bars. 
 
On the other hand, a possible source of systematic errors 
in the field-theoretic approach to 
critical phenomena could be the presence of non-analyticities on the real axis 
in
the $\beta$-function. Consider the critical behaviour of the coupling
constant and of the mass of a generic system. If one includes, beside the 
dominant correction to scaling with exponent
$\Delta_1$, generic subleading ones with exponents $2\Delta_1$,
$\Delta_2$ and analytic corrections, etc., one finds for the
$\beta$-function
\begin{eqnarray}
\lefteqn{W\left(g \right)=\frac{M(T)}{\frac{dM}{dT}}\frac{dg}{dT}=\frac{
\Delta_1}{\nu}
[\left(g-g^* \right)+a\left(g-g^{*}\right)^2}\nonumber\\ 
&&\quad
  +b\left(g-g^{*}\right)^{\frac{\Delta_2}{\Delta_1}}+c\left(g-g^{*}\right)^{
     \frac{1}{\Delta_1}}+\cdots ] \; .
\end{eqnarray}
One can easily see that confluent singularities on the real axis arise 
naturally if no specific hypothesis on the form of the corrections to scaling 
is made.
In this situation the usual summation method \cite{LGZJ} based on a Borel
transformation and a complex-plane mapping, in which the request of analyticity 
of
the $\beta$-function on the real axis plays a crucial role, will converge
very slowly.

Nickel \cite{Nickel} suggested a new type of analysis, consisting in fitting 
the 
$\beta$-function with functions with the expected cut singularity
on the real axis, such as hypergeometric
functions. With this type of analysis one finds \cite{Nickel2},
for the $O(n)$ $\sigma$-model
analytically continued to $n=0$,
that the zero of the $\beta$-function is at $g^{*}\approx1.39$, 
which is a value sensibly lower than the one obtained using the usual type
of analysis, $g^{*}=1.421\pm 0.008$. 
This fact reflects itself on the estimate of the critical
exponent $\gamma$ in the following way:
\begin{enumerate}
\item
the well-known value given by the standard analysis \cite{LGZJ} is
\begin{equation}
\gamma=1.1616 \pm 0.11  \left(g^* - 1.421 \right) \pm 0.0004
\end{equation}
\item
the one obtained with the analysis suggested by Nickel \cite{Nickel2} is
\begin{equation}
\gamma=1.1569 \pm 0.10  \left(g^* - 1.39 \right) \pm 0.0004
\end{equation}
\end{enumerate}

This  last value is in perfect agreement with our MC simulation \cite{noi}, 
as shown below.

\section{The Monte Carlo simulation}

One can hope to be able of discriminating between the two 
proposed types of analysis by means of a Monte Carlo simulation. 
The SAW is the best possible test-case because the
simulations are not affected by finite-size effects and 
because there are algorithms which have
autocorrelation times much smaller than those of the algorithms 
which are available for other
systems. To give an idea, for our algorithm the autocorrelation time in 
CPU units  scales as
$\tau\sim\xi ^2$, which can be compared with the best algorithms for 
the Ising model, for
which  $ \tau\sim\xi ^{d+z}$, with $z \approx 0.4$. 
A detailed description of the
algorithm used in this work can be found in \cite{joincut} . The algorithm 
works in
the ensemble of pairs of walks with fixed total length $N_{tot}$. One can make
inferences on the value of $\gamma$ from the observed distribution of length of 
one
of the two  walks, which is in our case
\begin{equation}
\pi (N_1)=\frac{c_{N_1}c_{N_{tot}-N_1}}{Z(N_{tot})} \; ,
\end{equation}
where $c_{N_1}$ denotes the number of walks of length $N_1$ and 
$Z=\sum_{N_1 = 0}^{N_{tot}} c_{N_1}c_{N_{tot}-N_1}$.

Instead of making  an {\em a priori} hypothesis on the form of the 
corrections to scaling to 
fit the MC data, we prefer to keep only the leading term in the scaling law for 
$c_N$
\begin{equation}
c_N \approx \mu ^N N^{\gamma -1}
\label{asymcN}
\end{equation}
and  make a type of analysis which is sensible to the presence of corrections 
to 
scaling, in the following way. We progressively reject those walks with length
shorter than a certain length $N_{min}$ and greater than $N_{tot}-N_{min}$ .  
Of course
the estimated values of $\gamma$, $\hat \gamma (N_{tot},N_{min})$,
depending on the range of allowed lengths and on $N_{tot}$, converge for
$N_{tot} \to
\infty$ and $N_{min} \to \infty$ to the correct value.

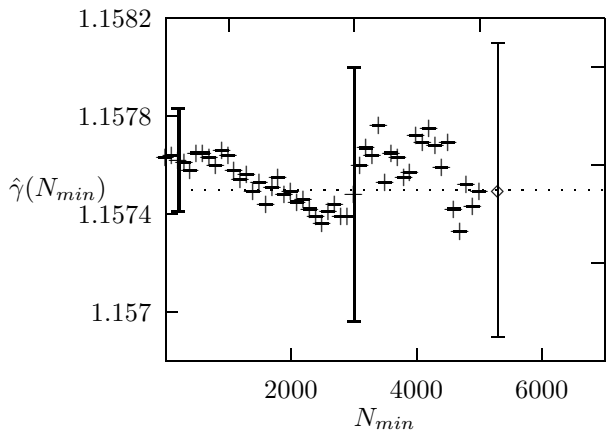
\begin{figure}[htb]
\vspace{9pt}

\setlength{\unitlength}{0.240900pt}
\ifx\plotpoint\undefined\newsavebox{\plotpoint}\fi
\sbox{\plotpoint}{\rule[-0.200pt]{0.400pt}{0.400pt}}%
\begin{picture}(974,675)(0,0)
\font\gnuplot=cmr10 at 10pt
\gnuplot
\sbox{\plotpoint}{\rule[-0.200pt]{0.400pt}{0.400pt}}%
\put(890.0,113.0){\rule[-0.200pt]{4.818pt}{0.400pt}}
\put(220.0,190.0){\rule[-0.200pt]{4.818pt}{0.400pt}}
\put(198,190){\makebox(0,0)[r]{$1.157$}}
\put(890.0,267.0){\rule[-0.200pt]{4.818pt}{0.400pt}}
\put(220.0,344.0){\rule[-0.200pt]{4.818pt}{0.400pt}}
\put(198,344){\makebox(0,0)[r]{$1.1574$}}
\put(890.0,421.0){\rule[-0.200pt]{4.818pt}{0.400pt}}
\put(220.0,498.0){\rule[-0.200pt]{4.818pt}{0.400pt}}
\put(198,498){\makebox(0,0)[r]{$1.1578$}}
\put(890.0,575.0){\rule[-0.200pt]{4.818pt}{0.400pt}}
\put(220.0,652.0){\rule[-0.200pt]{4.818pt}{0.400pt}}
\put(198,652){\makebox(0,0)[r]{$1.1582$}}
\put(319.0,632.0){\rule[-0.200pt]{0.400pt}{4.818pt}}
\put(417.0,113.0){\rule[-0.200pt]{0.400pt}{4.818pt}}
\put(417,68){\makebox(0,0){$2000$}}
\put(516.0,632.0){\rule[-0.200pt]{0.400pt}{4.818pt}}
\put(614.0,113.0){\rule[-0.200pt]{0.400pt}{4.818pt}}
\put(614,68){\makebox(0,0){$4000$}}
\put(713.0,632.0){\rule[-0.200pt]{0.400pt}{4.818pt}}
\put(811.0,113.0){\rule[-0.200pt]{0.400pt}{4.818pt}}
\put(811,68){\makebox(0,0){$6000$}}
\put(910.0,632.0){\rule[-0.200pt]{0.400pt}{4.818pt}}
\put(220.0,113.0){\rule[-0.200pt]{166.221pt}{0.400pt}}
\put(910.0,113.0){\rule[-0.200pt]{0.400pt}{129.845pt}}
\put(220.0,652.0){\rule[-0.200pt]{166.221pt}{0.400pt}}
\put(45,382){\makebox(0,0){ $\hat{\gamma}(N_{min})$}}
\put(565,23){\makebox(0,0){$N_{min}$}}
\put(220.0,113.0){\rule[-0.200pt]{0.400pt}{129.845pt}}
\put(742,382){\raisebox{-.8pt}{\makebox(0,0){$\diamond$}}}
\put(742.0,151.0){\rule[-0.200pt]{0.400pt}{111.296pt}}
\put(732.0,151.0){\rule[-0.200pt]{4.818pt}{0.400pt}}
\put(732.0,613.0){\rule[-0.200pt]{4.818pt}{0.400pt}}
\put(220,382){\usebox{\plotpoint}}
\put(220.00,382.00){\usebox{\plotpoint}}
\multiput(227,382)(20.756,0.000){0}{\usebox{\plotpoint}}
\put(240.76,382.00){\usebox{\plotpoint}}
\multiput(241,382)(20.756,0.000){0}{\usebox{\plotpoint}}
\multiput(248,382)(20.756,0.000){0}{\usebox{\plotpoint}}
\put(261.51,382.00){\usebox{\plotpoint}}
\multiput(262,382)(20.756,0.000){0}{\usebox{\plotpoint}}
\multiput(269,382)(20.756,0.000){0}{\usebox{\plotpoint}}
\put(282.27,382.00){\usebox{\plotpoint}}
\multiput(283,382)(20.756,0.000){0}{\usebox{\plotpoint}}
\multiput(290,382)(20.756,0.000){0}{\usebox{\plotpoint}}
\put(303.02,382.00){\usebox{\plotpoint}}
\multiput(304,382)(20.756,0.000){0}{\usebox{\plotpoint}}
\multiput(311,382)(20.756,0.000){0}{\usebox{\plotpoint}}
\put(323.78,382.00){\usebox{\plotpoint}}
\multiput(325,382)(20.756,0.000){0}{\usebox{\plotpoint}}
\multiput(332,382)(20.756,0.000){0}{\usebox{\plotpoint}}
\put(344.53,382.00){\usebox{\plotpoint}}
\multiput(345,382)(20.756,0.000){0}{\usebox{\plotpoint}}
\multiput(352,382)(20.756,0.000){0}{\usebox{\plotpoint}}
\put(365.29,382.00){\usebox{\plotpoint}}
\multiput(366,382)(20.756,0.000){0}{\usebox{\plotpoint}}
\multiput(373,382)(20.756,0.000){0}{\usebox{\plotpoint}}
\put(386.04,382.00){\usebox{\plotpoint}}
\multiput(387,382)(20.756,0.000){0}{\usebox{\plotpoint}}
\multiput(394,382)(20.756,0.000){0}{\usebox{\plotpoint}}
\put(406.80,382.00){\usebox{\plotpoint}}
\multiput(408,382)(20.756,0.000){0}{\usebox{\plotpoint}}
\multiput(415,382)(20.756,0.000){0}{\usebox{\plotpoint}}
\put(427.55,382.00){\usebox{\plotpoint}}
\multiput(429,382)(20.756,0.000){0}{\usebox{\plotpoint}}
\multiput(436,382)(20.756,0.000){0}{\usebox{\plotpoint}}
\put(448.31,382.00){\usebox{\plotpoint}}
\multiput(450,382)(20.756,0.000){0}{\usebox{\plotpoint}}
\multiput(457,382)(20.756,0.000){0}{\usebox{\plotpoint}}
\put(469.07,382.00){\usebox{\plotpoint}}
\multiput(471,382)(20.756,0.000){0}{\usebox{\plotpoint}}
\multiput(478,382)(20.756,0.000){0}{\usebox{\plotpoint}}
\put(489.82,382.00){\usebox{\plotpoint}}
\multiput(492,382)(20.756,0.000){0}{\usebox{\plotpoint}}
\multiput(499,382)(20.756,0.000){0}{\usebox{\plotpoint}}
\put(510.58,382.00){\usebox{\plotpoint}}
\multiput(513,382)(20.756,0.000){0}{\usebox{\plotpoint}}
\multiput(520,382)(20.756,0.000){0}{\usebox{\plotpoint}}
\put(531.33,382.00){\usebox{\plotpoint}}
\multiput(534,382)(20.756,0.000){0}{\usebox{\plotpoint}}
\multiput(541,382)(20.756,0.000){0}{\usebox{\plotpoint}}
\put(552.09,382.00){\usebox{\plotpoint}}
\multiput(555,382)(20.756,0.000){0}{\usebox{\plotpoint}}
\multiput(562,382)(20.756,0.000){0}{\usebox{\plotpoint}}
\put(572.84,382.00){\usebox{\plotpoint}}
\multiput(575,382)(20.756,0.000){0}{\usebox{\plotpoint}}
\multiput(582,382)(20.756,0.000){0}{\usebox{\plotpoint}}
\put(593.60,382.00){\usebox{\plotpoint}}
\multiput(596,382)(20.756,0.000){0}{\usebox{\plotpoint}}
\multiput(603,382)(20.756,0.000){0}{\usebox{\plotpoint}}
\put(614.35,382.00){\usebox{\plotpoint}}
\multiput(617,382)(20.756,0.000){0}{\usebox{\plotpoint}}
\multiput(624,382)(20.756,0.000){0}{\usebox{\plotpoint}}
\put(635.11,382.00){\usebox{\plotpoint}}
\multiput(638,382)(20.756,0.000){0}{\usebox{\plotpoint}}
\multiput(645,382)(20.756,0.000){0}{\usebox{\plotpoint}}
\put(655.87,382.00){\usebox{\plotpoint}}
\multiput(659,382)(20.756,0.000){0}{\usebox{\plotpoint}}
\multiput(666,382)(20.756,0.000){0}{\usebox{\plotpoint}}
\put(676.62,382.00){\usebox{\plotpoint}}
\multiput(680,382)(20.756,0.000){0}{\usebox{\plotpoint}}
\multiput(687,382)(20.756,0.000){0}{\usebox{\plotpoint}}
\put(697.38,382.00){\usebox{\plotpoint}}
\multiput(701,382)(20.756,0.000){0}{\usebox{\plotpoint}}
\multiput(708,382)(20.756,0.000){0}{\usebox{\plotpoint}}
\put(718.13,382.00){\usebox{\plotpoint}}
\multiput(722,382)(20.756,0.000){0}{\usebox{\plotpoint}}
\multiput(729,382)(20.756,0.000){0}{\usebox{\plotpoint}}
\put(738.89,382.00){\usebox{\plotpoint}}
\multiput(743,382)(20.756,0.000){0}{\usebox{\plotpoint}}
\multiput(750,382)(20.756,0.000){0}{\usebox{\plotpoint}}
\put(759.64,382.00){\usebox{\plotpoint}}
\multiput(764,382)(20.756,0.000){0}{\usebox{\plotpoint}}
\multiput(771,382)(20.756,0.000){0}{\usebox{\plotpoint}}
\put(780.40,382.00){\usebox{\plotpoint}}
\multiput(785,382)(20.756,0.000){0}{\usebox{\plotpoint}}
\multiput(792,382)(20.756,0.000){0}{\usebox{\plotpoint}}
\put(801.15,382.00){\usebox{\plotpoint}}
\multiput(805,382)(20.756,0.000){0}{\usebox{\plotpoint}}
\multiput(812,382)(20.756,0.000){0}{\usebox{\plotpoint}}
\put(821.91,382.00){\usebox{\plotpoint}}
\multiput(826,382)(20.756,0.000){0}{\usebox{\plotpoint}}
\multiput(833,382)(20.756,0.000){0}{\usebox{\plotpoint}}
\put(842.66,382.00){\usebox{\plotpoint}}
\multiput(847,382)(20.756,0.000){0}{\usebox{\plotpoint}}
\multiput(854,382)(20.756,0.000){0}{\usebox{\plotpoint}}
\put(863.42,382.00){\usebox{\plotpoint}}
\multiput(868,382)(20.756,0.000){0}{\usebox{\plotpoint}}
\multiput(875,382)(20.756,0.000){0}{\usebox{\plotpoint}}
\put(884.18,382.00){\usebox{\plotpoint}}
\multiput(889,382)(20.756,0.000){0}{\usebox{\plotpoint}}
\multiput(896,382)(20.756,0.000){0}{\usebox{\plotpoint}}
\put(904.93,382.00){\usebox{\plotpoint}}
\put(910,382){\usebox{\plotpoint}}
\sbox{\plotpoint}{\rule[-0.400pt]{0.800pt}{0.800pt}}%
\put(220,433){\makebox(0,0){$+$}}
\put(230,436){\makebox(0,0){$+$}}
\put(240,429){\makebox(0,0){$+$}}
\put(250,425){\makebox(0,0){$+$}}
\put(259,413){\makebox(0,0){$+$}}
\put(269,440){\makebox(0,0){$+$}}
\put(279,440){\makebox(0,0){$+$}}
\put(289,433){\makebox(0,0){$+$}}
\put(299,421){\makebox(0,0){$+$}}
\put(309,444){\makebox(0,0){$+$}}
\put(319,436){\makebox(0,0){$+$}}
\put(328,413){\makebox(0,0){$+$}}
\put(338,398){\makebox(0,0){$+$}}
\put(348,406){\makebox(0,0){$+$}}
\put(358,379){\makebox(0,0){$+$}}
\put(368,394){\makebox(0,0){$+$}}
\put(378,359){\makebox(0,0){$+$}}
\put(388,386){\makebox(0,0){$+$}}
\put(397,402){\makebox(0,0){$+$}}
\put(407,375){\makebox(0,0){$+$}}
\put(417,379){\makebox(0,0){$+$}}
\put(427,363){\makebox(0,0){$+$}}
\put(437,367){\makebox(0,0){$+$}}
\put(447,352){\makebox(0,0){$+$}}
\put(457,340){\makebox(0,0){$+$}}
\put(466,329){\makebox(0,0){$+$}}
\put(476,348){\makebox(0,0){$+$}}
\put(486,359){\makebox(0,0){$+$}}
\put(496,340){\makebox(0,0){$+$}}
\put(506,340){\makebox(0,0){$+$}}
\put(516,375){\makebox(0,0){$+$}}
\put(526,421){\makebox(0,0){$+$}}
\put(535,448){\makebox(0,0){$+$}}
\put(545,436){\makebox(0,0){$+$}}
\put(555,483){\makebox(0,0){$+$}}
\put(565,394){\makebox(0,0){$+$}}
\put(575,440){\makebox(0,0){$+$}}
\put(585,433){\makebox(0,0){$+$}}
\put(595,402){\makebox(0,0){$+$}}
\put(604,409){\makebox(0,0){$+$}}
\put(614,467){\makebox(0,0){$+$}}
\put(624,456){\makebox(0,0){$+$}}
\put(634,479){\makebox(0,0){$+$}}
\put(644,452){\makebox(0,0){$+$}}
\put(654,417){\makebox(0,0){$+$}}
\put(664,456){\makebox(0,0){$+$}}
\put(673,352){\makebox(0,0){$+$}}
\put(683,317){\makebox(0,0){$+$}}
\put(693,390){\makebox(0,0){$+$}}
\put(703,356){\makebox(0,0){$+$}}
\put(713,379){\makebox(0,0){$+$}}
\put(220,433){\usebox{\plotpoint}}
\put(210.0,433.0){\rule[-0.400pt]{4.818pt}{0.800pt}}
\put(210.0,433.0){\rule[-0.400pt]{4.818pt}{0.800pt}}
\put(230,436){\usebox{\plotpoint}}
\put(220.0,436.0){\rule[-0.400pt]{4.818pt}{0.800pt}}
\put(220.0,436.0){\rule[-0.400pt]{4.818pt}{0.800pt}}
\put(240.0,348.0){\rule[-0.400pt]{0.800pt}{39.026pt}}
\put(230.0,348.0){\rule[-0.400pt]{4.818pt}{0.800pt}}
\put(230.0,510.0){\rule[-0.400pt]{4.818pt}{0.800pt}}
\put(250,425){\usebox{\plotpoint}}
\put(240.0,425.0){\rule[-0.400pt]{4.818pt}{0.800pt}}
\put(240.0,425.0){\rule[-0.400pt]{4.818pt}{0.800pt}}
\put(259,413){\usebox{\plotpoint}}
\put(249.0,413.0){\rule[-0.400pt]{4.818pt}{0.800pt}}
\put(249.0,413.0){\rule[-0.400pt]{4.818pt}{0.800pt}}
\put(269,440){\usebox{\plotpoint}}
\put(259.0,440.0){\rule[-0.400pt]{4.818pt}{0.800pt}}
\put(259.0,440.0){\rule[-0.400pt]{4.818pt}{0.800pt}}
\put(279,440){\usebox{\plotpoint}}
\put(269.0,440.0){\rule[-0.400pt]{4.818pt}{0.800pt}}
\put(269.0,440.0){\rule[-0.400pt]{4.818pt}{0.800pt}}
\put(289,433){\usebox{\plotpoint}}
\put(279.0,433.0){\rule[-0.400pt]{4.818pt}{0.800pt}}
\put(279.0,433.0){\rule[-0.400pt]{4.818pt}{0.800pt}}
\put(299,421){\usebox{\plotpoint}}
\put(289.0,421.0){\rule[-0.400pt]{4.818pt}{0.800pt}}
\put(289.0,421.0){\rule[-0.400pt]{4.818pt}{0.800pt}}
\put(309,444){\usebox{\plotpoint}}
\put(299.0,444.0){\rule[-0.400pt]{4.818pt}{0.800pt}}
\put(299.0,444.0){\rule[-0.400pt]{4.818pt}{0.800pt}}
\put(319,436){\usebox{\plotpoint}}
\put(309.0,436.0){\rule[-0.400pt]{4.818pt}{0.800pt}}
\put(309.0,436.0){\rule[-0.400pt]{4.818pt}{0.800pt}}
\put(328,413){\usebox{\plotpoint}}
\put(318.0,413.0){\rule[-0.400pt]{4.818pt}{0.800pt}}
\put(318.0,413.0){\rule[-0.400pt]{4.818pt}{0.800pt}}
\put(338,398){\usebox{\plotpoint}}
\put(328.0,398.0){\rule[-0.400pt]{4.818pt}{0.800pt}}
\put(328.0,398.0){\rule[-0.400pt]{4.818pt}{0.800pt}}
\put(348,406){\usebox{\plotpoint}}
\put(338.0,406.0){\rule[-0.400pt]{4.818pt}{0.800pt}}
\put(338.0,406.0){\rule[-0.400pt]{4.818pt}{0.800pt}}
\put(358,379){\usebox{\plotpoint}}
\put(348.0,379.0){\rule[-0.400pt]{4.818pt}{0.800pt}}
\put(348.0,379.0){\rule[-0.400pt]{4.818pt}{0.800pt}}
\put(368,394){\usebox{\plotpoint}}
\put(358.0,394.0){\rule[-0.400pt]{4.818pt}{0.800pt}}
\put(358.0,394.0){\rule[-0.400pt]{4.818pt}{0.800pt}}
\put(378,359){\usebox{\plotpoint}}
\put(368.0,359.0){\rule[-0.400pt]{4.818pt}{0.800pt}}
\put(368.0,359.0){\rule[-0.400pt]{4.818pt}{0.800pt}}
\put(388,386){\usebox{\plotpoint}}
\put(378.0,386.0){\rule[-0.400pt]{4.818pt}{0.800pt}}
\put(378.0,386.0){\rule[-0.400pt]{4.818pt}{0.800pt}}
\put(397,402){\usebox{\plotpoint}}
\put(387.0,402.0){\rule[-0.400pt]{4.818pt}{0.800pt}}
\put(387.0,402.0){\rule[-0.400pt]{4.818pt}{0.800pt}}
\put(407,375){\usebox{\plotpoint}}
\put(397.0,375.0){\rule[-0.400pt]{4.818pt}{0.800pt}}
\put(397.0,375.0){\rule[-0.400pt]{4.818pt}{0.800pt}}
\put(417,379){\usebox{\plotpoint}}
\put(407.0,379.0){\rule[-0.400pt]{4.818pt}{0.800pt}}
\put(407.0,379.0){\rule[-0.400pt]{4.818pt}{0.800pt}}
\put(427,363){\usebox{\plotpoint}}
\put(417.0,363.0){\rule[-0.400pt]{4.818pt}{0.800pt}}
\put(417.0,363.0){\rule[-0.400pt]{4.818pt}{0.800pt}}
\put(437,367){\usebox{\plotpoint}}
\put(427.0,367.0){\rule[-0.400pt]{4.818pt}{0.800pt}}
\put(427.0,367.0){\rule[-0.400pt]{4.818pt}{0.800pt}}
\put(447,352){\usebox{\plotpoint}}
\put(437.0,352.0){\rule[-0.400pt]{4.818pt}{0.800pt}}
\put(437.0,352.0){\rule[-0.400pt]{4.818pt}{0.800pt}}
\put(457,340){\usebox{\plotpoint}}
\put(447.0,340.0){\rule[-0.400pt]{4.818pt}{0.800pt}}
\put(447.0,340.0){\rule[-0.400pt]{4.818pt}{0.800pt}}
\put(466,329){\usebox{\plotpoint}}
\put(456.0,329.0){\rule[-0.400pt]{4.818pt}{0.800pt}}
\put(456.0,329.0){\rule[-0.400pt]{4.818pt}{0.800pt}}
\put(476,348){\usebox{\plotpoint}}
\put(466.0,348.0){\rule[-0.400pt]{4.818pt}{0.800pt}}
\put(466.0,348.0){\rule[-0.400pt]{4.818pt}{0.800pt}}
\put(486,359){\usebox{\plotpoint}}
\put(476.0,359.0){\rule[-0.400pt]{4.818pt}{0.800pt}}
\put(476.0,359.0){\rule[-0.400pt]{4.818pt}{0.800pt}}
\put(496,340){\usebox{\plotpoint}}
\put(486.0,340.0){\rule[-0.400pt]{4.818pt}{0.800pt}}
\put(486.0,340.0){\rule[-0.400pt]{4.818pt}{0.800pt}}
\put(506,340){\usebox{\plotpoint}}
\put(496.0,340.0){\rule[-0.400pt]{4.818pt}{0.800pt}}
\put(496.0,340.0){\rule[-0.400pt]{4.818pt}{0.800pt}}
\put(516.0,175.0){\rule[-0.400pt]{0.800pt}{96.360pt}}
\put(506.0,175.0){\rule[-0.400pt]{4.818pt}{0.800pt}}
\put(506.0,575.0){\rule[-0.400pt]{4.818pt}{0.800pt}}
\put(526,421){\usebox{\plotpoint}}
\put(516.0,421.0){\rule[-0.400pt]{4.818pt}{0.800pt}}
\put(516.0,421.0){\rule[-0.400pt]{4.818pt}{0.800pt}}
\put(535,448){\usebox{\plotpoint}}
\put(525.0,448.0){\rule[-0.400pt]{4.818pt}{0.800pt}}
\put(525.0,448.0){\rule[-0.400pt]{4.818pt}{0.800pt}}
\put(545,436){\usebox{\plotpoint}}
\put(535.0,436.0){\rule[-0.400pt]{4.818pt}{0.800pt}}
\put(535.0,436.0){\rule[-0.400pt]{4.818pt}{0.800pt}}
\put(555,483){\usebox{\plotpoint}}
\put(545.0,483.0){\rule[-0.400pt]{4.818pt}{0.800pt}}
\put(545.0,483.0){\rule[-0.400pt]{4.818pt}{0.800pt}}
\put(565,394){\usebox{\plotpoint}}
\put(555.0,394.0){\rule[-0.400pt]{4.818pt}{0.800pt}}
\put(555.0,394.0){\rule[-0.400pt]{4.818pt}{0.800pt}}
\put(575,440){\usebox{\plotpoint}}
\put(565.0,440.0){\rule[-0.400pt]{4.818pt}{0.800pt}}
\put(565.0,440.0){\rule[-0.400pt]{4.818pt}{0.800pt}}
\put(585,433){\usebox{\plotpoint}}
\put(575.0,433.0){\rule[-0.400pt]{4.818pt}{0.800pt}}
\put(575.0,433.0){\rule[-0.400pt]{4.818pt}{0.800pt}}
\put(595,402){\usebox{\plotpoint}}
\put(585.0,402.0){\rule[-0.400pt]{4.818pt}{0.800pt}}
\put(585.0,402.0){\rule[-0.400pt]{4.818pt}{0.800pt}}
\put(604,409){\usebox{\plotpoint}}
\put(594.0,409.0){\rule[-0.400pt]{4.818pt}{0.800pt}}
\put(594.0,409.0){\rule[-0.400pt]{4.818pt}{0.800pt}}
\put(614,467){\usebox{\plotpoint}}
\put(604.0,467.0){\rule[-0.400pt]{4.818pt}{0.800pt}}
\put(604.0,467.0){\rule[-0.400pt]{4.818pt}{0.800pt}}
\put(624,456){\usebox{\plotpoint}}
\put(614.0,456.0){\rule[-0.400pt]{4.818pt}{0.800pt}}
\put(614.0,456.0){\rule[-0.400pt]{4.818pt}{0.800pt}}
\put(634,479){\usebox{\plotpoint}}
\put(624.0,479.0){\rule[-0.400pt]{4.818pt}{0.800pt}}
\put(624.0,479.0){\rule[-0.400pt]{4.818pt}{0.800pt}}
\put(644,452){\usebox{\plotpoint}}
\put(634.0,452.0){\rule[-0.400pt]{4.818pt}{0.800pt}}
\put(634.0,452.0){\rule[-0.400pt]{4.818pt}{0.800pt}}
\put(654,417){\usebox{\plotpoint}}
\put(644.0,417.0){\rule[-0.400pt]{4.818pt}{0.800pt}}
\put(644.0,417.0){\rule[-0.400pt]{4.818pt}{0.800pt}}
\put(664,456){\usebox{\plotpoint}}
\put(654.0,456.0){\rule[-0.400pt]{4.818pt}{0.800pt}}
\put(654.0,456.0){\rule[-0.400pt]{4.818pt}{0.800pt}}
\put(673,352){\usebox{\plotpoint}}
\put(663.0,352.0){\rule[-0.400pt]{4.818pt}{0.800pt}}
\put(663.0,352.0){\rule[-0.400pt]{4.818pt}{0.800pt}}
\put(683,317){\usebox{\plotpoint}}
\put(673.0,317.0){\rule[-0.400pt]{4.818pt}{0.800pt}}
\put(673.0,317.0){\rule[-0.400pt]{4.818pt}{0.800pt}}
\put(693,390){\usebox{\plotpoint}}
\put(683.0,390.0){\rule[-0.400pt]{4.818pt}{0.800pt}}
\put(683.0,390.0){\rule[-0.400pt]{4.818pt}{0.800pt}}
\put(703,356){\usebox{\plotpoint}}
\put(693.0,356.0){\rule[-0.400pt]{4.818pt}{0.800pt}}
\put(693.0,356.0){\rule[-0.400pt]{4.818pt}{0.800pt}}
\put(713,379){\usebox{\plotpoint}}
\put(703.0,379.0){\rule[-0.400pt]{4.818pt}{0.800pt}}
\put(703.0,379.0){\rule[-0.400pt]{4.818pt}{0.800pt}}
\end{picture}
\caption{Estimates of $\gamma$ obtained from the 
weighted average of the results
with $N_{tot}=20000$ (number of iterations $N_{iter}=10^8$) and $N_{tot}=40000$
($N_{iter}=8.5 \cdot 10^8$). For clarity the statistical error is reported 
only in two cases. The dotted line and the rightmost point
indicated by a diamond is our final estimate with its 
error bar.} 
\label{stimafinale}
\end{figure}

High-statistics runs at different total length, giving an idea of how strong 
the
effect of correction to scaling can be, are shown in Table~\ref{tab:tavola}. 
\begin{table*}[t]
\setlength{\tabcolsep}{1.5pc}
\newlength{\digitwidth} \settowidth{\digitwidth}{\rm 0}
\catcode`?=\active \def?{\kern\digitwidth}
\caption{Estimates of $\gamma$ for $N_{tot}=200$ and $N_{tot}=2000$.
$N_{iter}$ is the number of iterations.}
\label{tab:tavola}
\begin{tabular*}{\textwidth}{@{}l@{\extracolsep{\fill}}rrrr}
\hline
    \multicolumn{2}{c}{$N_{tot} = 200$} 
   & \multicolumn{2}{c}{$N_{tot} = 2000$} \\
\hline
    $N_{min}$  &  $\hat \gamma$  &
    $N_{min}$  &  $\hat \gamma$  \\
\hline
1  &  $ 1.15288 \pm 0.00011 $ &
1  &  $ 1.15782 \pm 0.00013 $ \\
10 &  $ 1.15808 \pm 0.00021 $ &
100 &  $ 1.15802 \pm 0.00028 $ \\
20 &   $ 1.15866 \pm 0.00034 $ &
200 &  $ 1.15811 \pm 0.00045 $ \\
30 &   $ 1.15875 \pm 0.00053 $ &
300 &  $ 1.15838 \pm 0.00071 $ \\
40 &  $ 1.15999 \pm 0.00084 $ &
400 &  $ 1.1598\hphantom{0} \pm 0.0011\hphantom{0} $ \\
50 &   $ 1.1605\hphantom{0} \pm 0.0014\hphantom{0} $ &
500 &  $ 1.1584\hphantom{0} \pm 0.0019\hphantom{0} $ \\
\hline
    $N_{iter}$  &  $5 \cdot 10^8$  &
    $N_{iter}$  &  $6.2 \cdot 10^8$  \\
\hline

\end{tabular*}
\end{table*}
The run at $N_{tot}=200$ shows a strong
dependence on the cut $N_{min}$, but is in agreement, for large values of the 
cut, with 
\cite{LGZJ} and with previous MC
simulations \cite{Grassberger}. For $N_{tot}=2000$, the estimated values are
lower and much flatter, but still sensibly biased by corrections to scaling.
For the final estimate we used a weighted average of the estimates 
obtained using walks 
of length $N_{tot}=20000$ and $N_{tot}=40000$. 
The effective exponents are shown in Figure~\ref{stimafinale}, 
where the
last point on the right denotes the final estimate with its error bar. 
The estimates reported in Fig. \ref{stimafinale} show a very small 
dependence on $N_{min}$ but, of course, one cannot conclude that
the estimates are not biased by corrections-to-scaling 
effects. Indeed to understand the residual systematic effects 
we must compare estimates with different values of $N_{tot}$. 
Assuming that the corrections to Eq. (\ref{asymcN})  scale with a 
subleading exponent $\Delta\gtapprox 0.5$, one confirms that our data
at the larger values of $N_{tot}$ have a systematic bias which is 
less than the statistical error.
A very conservative way of determining the residual systematic error consists in
making  the hypothesis that the effective estimates $\hat{\gamma}$ tend toward
 the true value following the law
\begin{equation}
\hat{\gamma}=\gamma + \frac{B}{{N_{tot}}^{\Delta}} \; .
\label{syst}
\end{equation}
Fitting our data to this law, requiring that~(\ref{syst}) 
reproduces the highest
estimated exponent  for ${N_{tot}}=2000$ and the lowest one for
${N_{tot}}=40000$ with $\Delta = 0.5$, we estimate $B \approx 0.06$.
This result clearly leads to an overestimate of the systematic error if $\Delta
> 0.5$.
With this type of analysis of the data, we estimate
\begin{equation}
\gamma = 1.1575 \pm 0.0006 \; .
\end{equation}

\section{Conclusions}
The present work indicates that the previous estimate of the critical exponent 
$\gamma$ obtained from the
RG $g$-expansion is significantly biased upward, possibly because of 
the presence of strong confluent
singularities of the $\beta$-function on the real axis at $g^*$. We hope that
further analytic work can be done in this direction to clarify this point.
Previous Monte Carlo and exact enumeration determinations appear to 
be incorrect as well, showing how important and difficult is 
the determination of the systematic error due to 
corrections to scaling.

\end{document}